\documentclass[%
 preprint,
nofootinbib,
 amsmath,amssymb,
 aps,
]{revtex4}

\usepackage{graphicx}
\usepackage{dcolumn}
\usepackage{bm}

\usepackage{braket}
\usepackage{mathrsfs}

\newcommand{\Omegazero}{\Omega^{(0)}}
\newcommand{\dd}{\mathrm{d}}
\newcommand{\angstrom}{\mbox{\normalfont\AA}}
\newcommand{\um}[1]{\ensuremath{\mathrm{\,#1}}} 
\renewcommand{\vec}{\bm}

\begin{document}

\preprint{}

\title{Fluctuation relations for dissipative systems in constant external magnetic field: theory and molecular dynamics simulations}

\author{Alessandro Coretti}
\affiliation{ORCID: 0000-0002-7131-3210}
\affiliation{%
 Department of Mathematical Sciences, Politecnico di Torino, Corso Duca degli Abruzzi 24, I-10129 Torino, Italy
}%
\affiliation{
Centre Europ\'een de Calcul Atomique et Mol\'eculaire (CECAM), \'Ecole Polytechnique F\'ed\'erale de Lausanne, Batochime, Avenue Forel 2, 1015 Lausanne, Switzerland
}%
\affiliation{
Istituto Nazionale di Fisica Nucleare, Sezione di Torino, Via P. Giura 1, I-10125 Torino, Italy
}%

\author{Lamberto Rondoni}
\affiliation{ORCID: 0000-0002-4223-6279}
\affiliation{%
Department of Mathematical Sciences, Politecnico di Torino, Corso Duca degli Abruzzi 24, I-10129 Torino, Italy%
}%
\affiliation{
Istituto Nazionale di Fisica Nucleare, Sezione di Torino, Via P. Giura 1, I-10125 Torino, Italy
}%

\author{Sara Bonella}
\email{sara.bonella@epfl.ch}
\affiliation{ORCID: 0000-0003-4131-2513 \\
Centre Europ\'een de Calcul Atomique et Mol\'eculaire (CECAM), \'Ecole Polytechnique F\'ed\'erale de Lausanne, Batochime, Avenue Forel 2, 1015 Lausanne, Switzerland
}%

\date{\today}

\begin{abstract}
It has recently been pointed out that Hamiltonian particle systems in constant magnetic fields satisfy generalized time-reversal symmetries that enable to prove useful statistical relationships based on equilibrium phase-space probability distributions without the need to invert, as commonly considered necessary, the magnetic field. Among these relations, that hold without need of Casimir modifications, one finds the standard linear response Green-Kubo relations, and consequently the Onsager reciprocal relations. Going beyond linear response is also possible, for instance in terms of transient and steady state Fluctuation Relations (FRs). Here we highlight how the generalized time-reversal symmetries ensure that the (transient) FRs theory directly applies also for systems in external magnetic fields. Furthermore we show that transient FR can indeed be verified in nonequilibrium molecular dynamics simulations, for systems subjected to magnetic and electric fields, which are thermostatted \emph{\`a la} Nos\'e-Hoover. The result is nontrivial because, since it is not immediate within which sizes and time scales the effects can actually be observable, it is not obvious what one may obtain by real molecular dynamics simulations.
\end{abstract}

\maketitle

\section{Introduction}
It is well known that the evolution equations\footnote{In this work, we focus on classical systems.} of charged systems subject to an external magnetic field are not invariant under the standard time-reversal transformation defined \emph{via} inversion of the momenta
\begin{equation}
\mathcal{M}_s (\vec{r},\vec{p}) = (\vec{r},-\vec{p}) ~, \quad \forall (\vec{r}, \vec{p}) \doteq \Gamma \in  \mathfrak{M}
\label{eq:oldsym}
\end{equation}
combined with the change $t\rightarrow - t$. (In the equation above, $\Gamma$ is a point in the phase space $\mathfrak{M}$ of an $N$-particle system, with positions $\vec{r} = \{\vec{r}_i\}_{i=1}^{N}$ and momenta $\vec{p} = \{\vec{p}_i\}_{i=1}^{N}$.) This fact originated the idea that these systems require special treatment when discussing properties based on time reversibility. In particular, because the currents (and therefore the magnetic field that they generate) are reversed under $\mathcal{M}_s$, classical text books~\cite{landau:1980-book,degroot:1984-book} as well as current literature~\cite{barbier:2018a} present statistical relationships in the presence of a magnetic field using pairs of systems with identical interparticle interactions but under magnetic fields of opposite signs. For example,
the Onsager reciprocal relations were adapted by Casimir to relate cross-transport coefficients of systems under opposite magnetic fields~\cite{casimir:1945}. Similarly,
Kubo~\cite{kubo:1966} derived symmetry properties of time-correlation functions of two such systems subject to $\vec{B}$ and $- \vec{B}$. In nonequilibrium statistical mechanics, results for currents, response, and Fluctuation Relations (FRs) are typically presented under the same conditions~\cite{gaspard:2013,barbier:2018a,wang:2014,saito:2008}. All these results are to be contrasted with their ``standard'' counterparts that refer to a single system. 
 
 The situation described above is somewhat unsatisfactory for two main reasons. The first is conceptual: the introduction of a second system, while physically correct, blurs the distinction between the system and its external environment. Indeed, in the evolution equations of the system, the magnetic field appears as an external agent whose physical origin (\emph{e.g.} moving charges originating a current) is not associated to active degrees of freedom in the dynamical system. Its inversion then implicitly implies an extension of the system to include the sources of the magnetic field, applying $\mathcal{M}_s$ to the extended system, and then forgetting again about the additional degrees of freedom. The second reason is practical: this commonly adopted approach reduces the predictive power of the statistical relationships mentioned above. For example, within linear response theory, null values of transport coefficients in experiments concerning a single system in a given magnetic field cannot be predicted based on symmetry properties of the time-correlation functions. Similarly, in the context of nonlinear response, which includes the fluctuation relations~\cite{evans:1993,gallavotti:1995b,evans:2002,evans:2005,rondoni:2007}, null cumulants of the dissipation cannot be identified \emph{via} symmetry \cite{barbier:2018a}.

Recently however, it was demonstrated that, for systems in a constant external magnetic field, these difficulties can be overcome, recovering the full predictive power of statistical mechanics~\cite{bonella:2014}. The starting observation for these recent developments is that invariance of the Hamiltonian (and hence of the dynamical system) under Eq.~\eqref{eq:oldsym} is a sufficient but not necessary condition for establishing the properties mentioned above. Following a known approach in nonequilibrium statistical mechanics, alternative time-reversal operators --- that do not necessitate inversion of the magnetic field --- can be introduced~\cite{roberts:1992,bonella:2017a,coretti:2018a,carbone:2020} and used instead of $\mathcal{M}_s$ to reinstate standard proofs. These new symmetries lack the intuitive property of retracing the coordinates in the backward propagation in pairs of trajectories with opposite momenta, but they nonetheless identify pairs of trajectories with opposite values of relevant observables (\emph{e.g.} elements of the diffusion tensor, instantaneous dissipation) and their physical effects can be predicted and measured. This was illustrated numerically for the case of time-correlation functions in Refs.~\cite{bonella:2017a,coretti:2018a} and for fluctuation relations in the presence of orthogonal electric and magnetic fields in Ref.~\cite{coretti:2020b}. The lack of experimental evidence of the violation of the Onsager reciprocal relations~\cite{luo:2020} might also be explained \emph{via} these generalized time-reversal operators. 

In this paper we consider again the nonequilibrium behavior of classical charged systems in external magnetic and electric fields and generalize and consolidate previous work. In particular, we consider the case of parallel fields in which net currents and dissipation arise: the situation for which FRs have been developed~\cite{evans:2002,rondoni:2007}. We build on two generalized time-reversal operators introduced in Ref.~\cite{coretti:2018a} to motivate the validity of a transient FR that benefits --- contrary to the existing literature --- from a fully single-system (single magnetic field) description. The analytical results are then illustrated by molecular dynamics simulations of a realistic model of NaCl. The evolution equations are integrated \emph{via} a symplectic and time-reversible algorithm that includes a modified Nos\'e-Hoover thermostat to enforce constant temperature~\cite{mouhat:2013}. These simulations show the odd parity of the dissipation under the proposed generalized time-reversal operators and verify the validity of the transient fluctuation relation for a representative value of the electric field.

\section{Theory}
Let us  consider $N$ particles of charge $q_i$ and mass $m_i$ in three dimensions and in the presence of external uniform and static electric and magnetic fields. The Hamiltonian of the system is
\begin{equation}
\begin{aligned}
\label{eq:EM_Ham}
H(\Gamma) &= H_0(\Gamma) - \sum_{i=1}^Nq_i\vec{E}\cdot\vec{r}_i = \\
&= \sum_{i=1}^N\frac{\bigl(\vec{p}_i - q_i\vec{A}(\vec{r}_i)\bigr)^2}{2m_i} + \sum_{i,j<i}^NV(r_{ij}) - \sum_{i=1}^Nq_i\vec{E}\cdot\vec{r}_i
\end{aligned}
\end{equation}
In the equation above, $\vec{A}(\vec{r})$ is the vector potential associated to the magnetic field $\vec{B} = \vec{\nabla}_{\vec{r}} \times \vec{A}(\vec{r})$, $\vec{E}$ is the electric field and $V(r_{ij})$ a pairwise additive interaction potential, depending only on the modulus of the distance between particles: $r_{ij} = |\vec{r}_i - \vec{r}_j|$. We set $\vec{E} = (0, 0, E_z)$ and $\vec{B} = (0, 0, B_z)$, \emph{i.e.} the fields are parallel and oriented along the $z$-axis. In the Coulomb gauge ($\vec{\nabla}_{\vec{r}} \cdot \vec{A}(\vec{r}) = 0$), a valid choice for the vector potential is $\vec{A}(\vec{r}) = B_z/2(-y, x, 0)$.~\footnote{The choice of the gauge does not affect the discussion below since it cannot affect the evolution equations, see also~\cite{carbone:2020}.} This setting, while not completely general, is typically adopted to discuss the time-reversal properties of systems in external magnetic fields~\cite{gaspard:2013,barbier:2018a,jayannavar:2007,poria:2016} and it describes relevant physical situations. In particular, this orientation of the electric and magnetic fields ensures the presence of dissipation in the system and this is the framework in which FRs and their corollaries are generally considered. 
To proceed, we couple the system to the Nos\'e-Hoover thermostat. This thermostat is commonly adopted in molecular dynamics simulations and, at equilibrium, it provides a sampling of the canonical ensemble. In particular, we shall consider a modified version of this thermostat that was introduced to account for the presence of magnetic and electric fields~\cite{mouhat:2013}. The corresponding dynamical system is
\begin{equation}
\label{eq:NH-EoM}
\begin{aligned}
\frac{\dd x_i}{\dd t} &= \frac{p^x_i}{m_i} + \omega_iy_i \\
\frac{\dd y_i}{\dd t} &= \frac{p^y_i}{m_i} - \omega_ix_i \\
\frac{\dd z_i}{\dd t} &= \frac{p^z_i}{m_i} \\
\frac{\dd \ln s}{\dd t} & = \xi
\end{aligned}
\qquad\qquad
\begin{aligned}
\frac{\dd p^x_i}{\dd t} &= F^x_i + \omega_i(p^y_i - m_i\omega_ix_i)  - \xi(p^x_i + m_i\omega_iy_i)\\
\frac{\dd p^y_i}{\dd t} &= F^y_i - \omega_i(p^x_i + m_i\omega_iy_i) - \xi(p^y_i - m_i\omega_ix_i)\\
\frac{\dd p^z_i}{\dd t} &= F^z_i + q_iE_z - \xi p^z_i \\
\frac{\dd \xi}{\dd t} &= \frac{1}{\tau^2_{\mathrm{NH}}}\biggl[\frac{K(\Gamma) - K^*}{K^*}\biggr] = \frac{\delta K(\Gamma)}{\tau^2_{\mathrm{NH}}}
\end{aligned}
\end{equation}
where $\tau_{\mathrm{NH}}$ is the characteristic time of the thermostat and $K(\Gamma)$ is the instantaneous kinetic energy of the physical degrees of freedom. This kinetic energy fluctuates around the target value $K^*$, which is related to the temperature, $T$, of the system \emph{via} $\beta = G/2K^*$ ($\beta=1/k_BT$, $k_B$ the Boltzmann constant and $G$ are the degrees of freedom of the system). As hinted by the notation in Eq.~(\ref{eq:EM_Ham}), in the following we shall consider the effect of the electric field on the system at equilibrium in the presence of the external magnetic field. In Ref.~\cite{mouhat:2013} it was shown that the dynamical system \eqref{eq:NH-EoM} conserves the quantity $H_{\mathrm{NH}}(\Gamma, \xi, s) = H_0(\Gamma) + K^*\bigl[\tau^2_{\mathrm{NH}}\xi^2 + 2\ln s\bigr]$. When no electric field is present, the dynamics samples the equilibrium distribution
\begin{equation}\label{eq:ExtCanonicalDens}
f_0(X) = \mathcal{Z}^{-1}\exp[-\beta H_0(\Gamma)]\exp\biggl[-\frac{G\tau^2_{\mathrm{NH}}\xi^2}{2}\biggr]
\end{equation}
where $\mathcal{Z}$ is the partition function 
and $X$ denotes a point in the extended phase space of the system, with $X = (\Gamma, \xi)$. As in standard Nos\'e-Hoover dynamics, the marginal probability obtained integrating Eq.~\eqref{eq:ExtCanonicalDens} with respect to $\xi$ is the canonical density, in magnetic field, for the physical variables. 

Let us now discuss the behavior under time reversal of this dynamical systems. Direct inspection shows that, as expected, standard time reversal does not hold even considering a natural extension $\mathcal{M}_s^{\mathrm{ext}}$ which includes the Nos\'e-Hoover auxiliary variables by leaving $s$ unchanged and changing the sign of $\xi$. This is due to the coupling between coordinates and momenta induced by the magnetic field and seems to imply that a standard treatment of equilibrium and nonequilibrium relationships based on time-reversal is indeed impossible. However, the proof of these relationships requires the existence of (at least) \textbf{one} valid time-reversal operator and the even parity of the equilibrium distribution under this operator, but it does not prescribe the specific form of the operator and, in particular, it does not fix it to $\mathcal{M}_s$ or $\mathcal{M}^{\mathrm{ext}}_s$. In fact, generalized time-reversal operators, different from $\mathcal{M}_s$, have already appeared in the literature to investigate the equilibrium and nonequilibrium statistical mechanics of molecular dynamics systems, \emph{i.e.} deterministic particle systems~\cite{EMbook}. To adapt these definitions to our problem, let us denote as $\mathfrak{M}_{\mathrm{ext}}$ the extended phase space spanned by the dynamical system~\eqref{eq:NH-EoM}, and as $\mathcal{U}_t$ the associated time-evolution operator. Generalized time-reversal operators in this extended phase space are defined, in complete analogy with what is done in the physical phase space, as involutions $\mathcal{M}_{\mathrm{ext}}$ satisfying
\begin{equation}
\label{eq:TRI}
\mathcal{U}_{-t} X = \mathcal{M}_{\mathrm{ext}} \mathcal{U}_t \mathcal{M}_{\mathrm{ext}} X \quad
\forall t \in \mathbb{R} \ , \ \forall X \in \mathfrak{M}_{\mathrm{ext}}
\end{equation}
Crucially for the developments presented in the following, two such operators can be defined for~\eqref{eq:NH-EoM}: the dynamical system is in fact invariant under
\begin{subequations}
\label{eq:NH_TRS}
\begin{align}
\mathcal{M}_{\mathrm{ext}}^{(3)}(\Gamma, s, \xi) &= (-x,y,z,p^x,-p^y,-p^z,s,-\xi) \\
\mathcal{M}_{\mathrm{ext}}^{(4)}(\Gamma, s, \xi) &= (x,-y,z,-p^x,p^y,-p^z,s,-\xi) 
\end{align}
\end{subequations}
combined with time inversion, and both operators satisfy the definition Eq.~\eqref{eq:TRI}. The notation adopted above reflects the nomenclature introduced in Ref.~\cite{coretti:2018a} where related symmetries --- established in the absence of a thermostat --- were first introduced. Note that the equilibrium density Eq.~\eqref{eq:ExtCanonicalDens} is even under these transformations. As mentioned above, the validity of the time-reversal operators defined in Eq.~\eqref{eq:NH_TRS} and the even parity of the equilibrium probability density of the system is a sufficient condition to reinstate standard proofs of relevant statistical mechanics relationships. We stress that these new time-reversal symmetries touch only the active degrees of freedom in the dynamical system and do not require inversion of the magnetic field. Based on these symmetries, we can then derive interesting results within a \textbf{single-system} (single magnetic field) discussion of the dynamics. For example, following the derivation in Ref.~\cite{coretti:2018a}, it can be shown within linear response theory that the $yz$ and $xz$ components of the diffusion and conductivity tensors of the system must be zero. 

\begin{figure}[htb]
\centering
\includegraphics[width=\columnwidth]{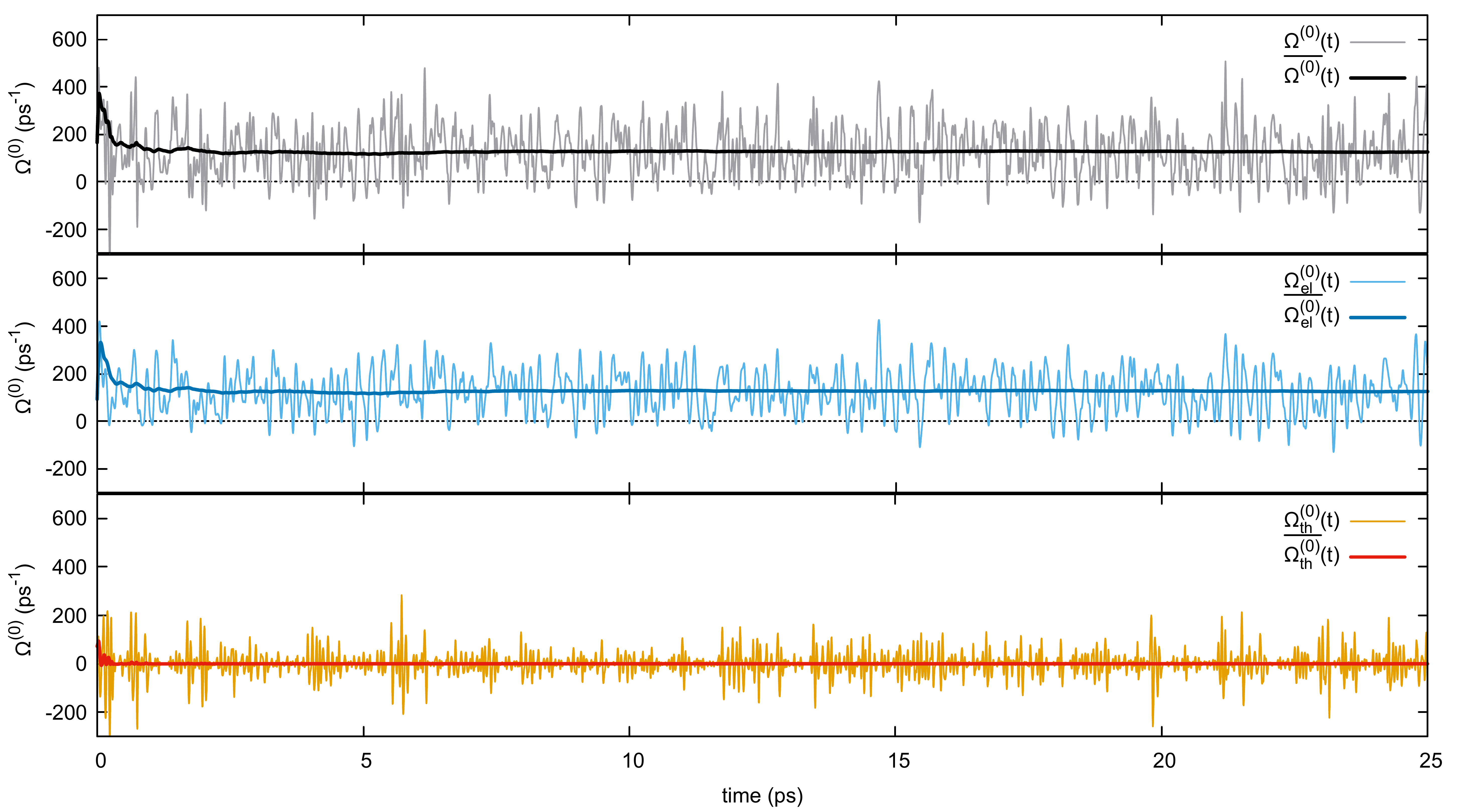}
\caption{Different contributions to the dissipation function $\Omegazero$ computed along 25\um{ps} of evolution. The contribution due to the thermostat $\Omegazero_{\mathrm{th}}$ is shown in the lower panel, the contribution due to the electric field $\Omegazero_{\mathrm{el}}$ is in the middle panel, while the upper panel shows the total dissipation.}
\label{fig:Omega_thermostat}
\end{figure}

In the following, we shall consider the implications of the newly introduced time-reversal operators on nonequilibrium properties of the system, focusing in particular on the transient fluctuation relation. The key quantity in this relation is the instantaneous dissipation function that, in the extended phase space, is defined as
\begin{equation}
\label{eq:omegazero_NH_def}
\Omegazero(X) \doteq - \dot{X} \cdot \nabla_X \ln f_0 - \Lambda(X) 
\end{equation} 
where $\Lambda(X)= \nabla_{X}\cdot\dot{X}$ is the phase-space expansion rate. Substitution of Eq.~\eqref{eq:ExtCanonicalDens} in the definition above shows, after some algebra~\cite{coretti:2020b}, that $\dot{X}\cdot \nabla_X \ln f_0 = G\xi - \beta \sum^N_{i=1}q_i\dot{\vec{r}}_i\cdot\vec{E} -G\xi\delta K(\Gamma)$. Furthermore, $\Lambda(X) = -G\xi$. Combining these results in Eq.~\eqref{eq:omegazero_NH_def} we obtain
\begin{equation}
\label{eq:omegazero_NH}
\Omegazero(X) = \mathscr{V} \beta \vec{J}(\Gamma)\cdot\vec{E} + G\xi\delta K(\Gamma)
\end{equation} 
where $\mathscr{V}$ is the volume of the system and $\vec{J}(\Gamma)=\mathscr{V}^{-1}\sum^N_{i=1}q_i\dot{\vec{r}}_i$ is the microscopic current. For the Nos\'e-Hoover system examined here then, the dissipation function is given by the sum of two contributions. The first one is the usual dissipative term related to the current induced by the electric field; the second accounts for the dissipation originating from the thermal gradient between the system and the reservoir --- represented in this case by the $(s,\xi)$ variables. The average dissipation over a time-leg $\tau$ is defined as
\begin{equation}
\label{eq:average_Om}
\overline{\Omegazero}_{0,\tau}(X) \doteq \frac{1}{\tau} \int_{0}^{\tau}\dd s\Omegazero(\mathcal{U}_s X)
\end{equation}
For $\tau \gg \tau_{\mathrm{NH}}$, it can be shown~\cite{coretti:2020b} that the contribution to this average due to the temperature gradient is negligible compared to the term expressing the dissipation associated with the currents driven by the electric field. This is illustrated in Figure~\ref{fig:Omega_thermostat}, where the dissipation function is analyzed for a 25\um{ps} molecular dynamics run of liquid NaCl with realistic interactions (the details of the simulation are provided in the next section). The bottom panel of the figure shows the instantaneous (orange curve) and average (red curve) thermal dissipation (second term on the right hand side of Eq.~\eqref{eq:omegazero_NH} and its time average, respectively). As it can be seen, the instantaneous thermal dissipation is finite, but it oscillates around zero so that it sums to a null contribution in the average. The electric dissipative term on the other hand, shown in the middle panel of Figure~\ref{fig:Omega_thermostat} (first term on the right hand side of Eq.~\eqref{eq:omegazero_NH} --- in cyan --- and its time average --- in blue ---, respectively), has nonzero instantaneous and average value and clearly dominates the total dissipation, upper panel of the figure. 
The analytical result in Ref.~\cite{coretti:2020b} and the numerical results presented here then support the fact that, for sufficiently long averaging windows, the dissipation assumes a form which does not depend on the specific thermostat adopted. A similar effect was observed in Ref.~\cite{evans:2005}, where oscillations of the interparticle potential energy had to average out, in order for the currents contributions to dominate the dissipation function, and for the steady-state FR to hold. In that case, at the small fields relevant for linear response, this was not guaranteed to be the case.

\begin{figure}[htb]
\centering
\includegraphics[width=\columnwidth]{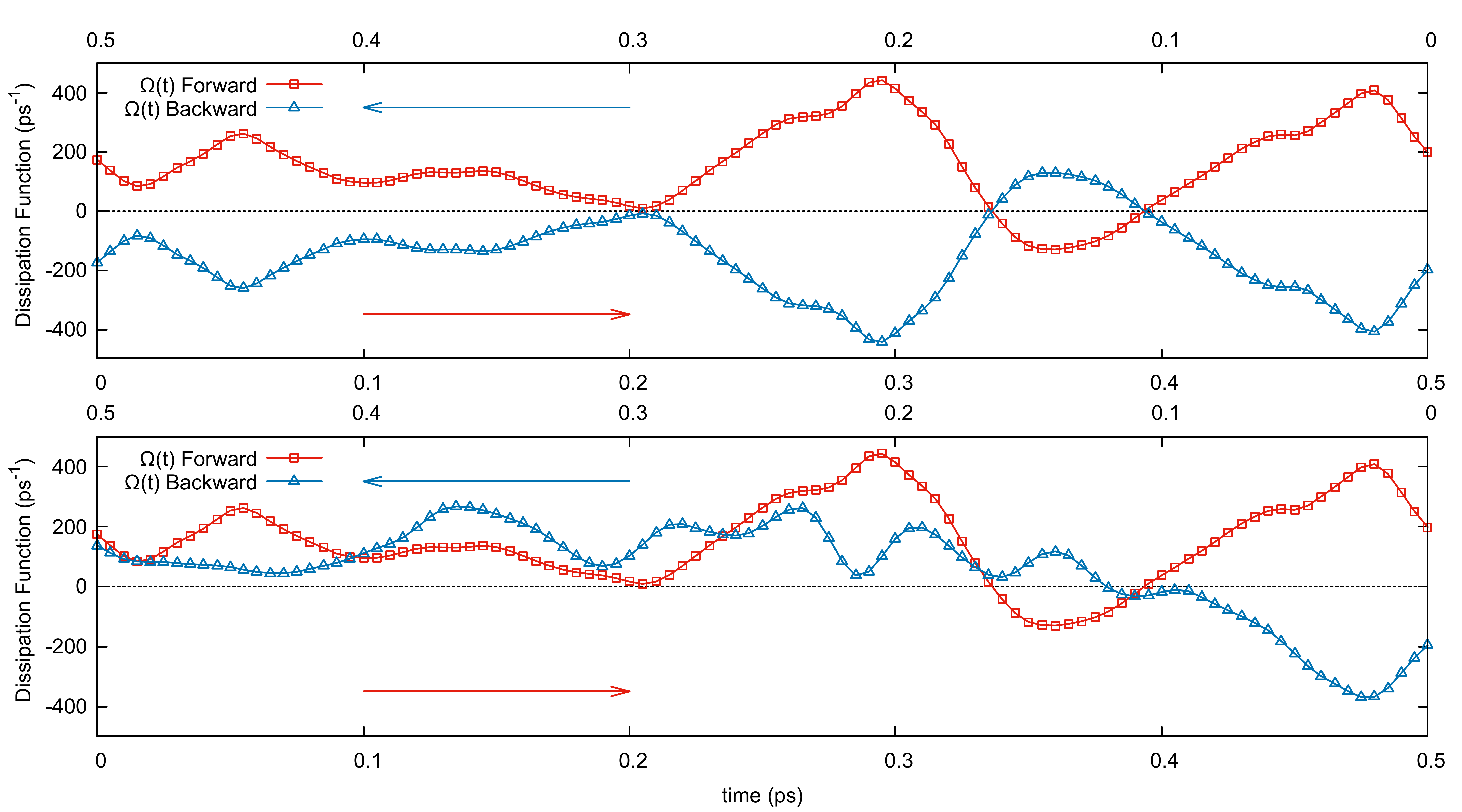}
\caption{Instantaneous dissipation function from Eq.~\eqref{eq:omegazero_NH_def} for 500\um{fs} of the ``forward'' evolution (red curve, open squares) and for 500\um{fs} of the ``backward'' trajectory (blue curve, open triangles) obtained \emph{via} $\mathcal{M}^{(3)}_{\mathrm{ext}}$ (upper panel). The opposite values of the dissipation demonstrate the odd signature under the generalized time-reversal transformation used. The same behavior is not observed when the backward trajectory is obtained applying $\mathcal{M}_s^{\mathrm{ext}}$ (bottom panel) as the presence of the magnetic field breaks the symmetry of the system under this transformation. Results are for the nonequilibrium simulation set-up described in Section~\ref{sec:results} for liquid NaCl.}
\label{fig:M3}
\end{figure}

Furthermore, as required, in this system the instantaneous dissipation is odd under Eqs.~\eqref{eq:NH_TRS}. In Figure~\ref{fig:M3} (top panel) we show the behavior of this quantity under $\mathcal{M}_{\mathrm{ext}}^{(3)}$ (the behavior under $\mathcal{M}_{\mathrm{ext}}^{(4)}$ is the same). In the figure, $\Omegazero(t)$ is computed along a ``forward'' trajectory (in red), and along the ``backward'' trajectory (in blue) identified by $\mathcal{M}_{\mathrm{ext}}^{(3)}$ for the same liquid NaCl simulation previously considered. More in detail, the two curves are obtained as follows: the dynamical system~\eqref{eq:NH-EoM} is evolved for $500\um{fs}$ (``forward'' trajectory) and $\Omegazero(t)$ is computed along the trajectory. The operator $\mathcal{M}_{\mathrm{ext}}^{(3)}$ is then applied to the phase-space point obtained at the end of the evolution and the system is evolved again \emph{via} Eqs.~\eqref{eq:NH-EoM} for $500\um{fs}$ starting from the transformed point (``backward'' trajectory). Along this trajectory we compute again $\Omegazero(t)$. The odd parity of the dissipation is apparent from the figure. As a curiosity, in the bottom panel of Figure~\ref{fig:M3}, we show the results for calculations in which the ``backward'' trajectory corresponds to standard time-reversal in the extended phase space. The figure clearly shows the lack of a specific signature for the dissipation under this symmetry.
Having computed and characterized the dissipation function, we now move to the discussion of the associated transient fluctuation relation. This relation provides, in fact, an explicit expression for the ratio of the initial probabilities to find the average dissipation function, $\overline{\Omegazero}_{0,\tau}$ in a neighborhood of size $\delta$ of the values $A$ and of $-A$. Defining the subset of the phase space where the average dissipation takes values in the interval $(\pm A)_{\delta} = (\pm A-\delta, \pm A+\delta)$ by $\{\overline{\Omegazero}_{0,\tau}\}_{(\pm A)_{\delta}}$, this expression is given by~\cite{searles:2007,rondoni:2007}
\begin{equation}
\label{eq:t-FR-fin}
\begin{aligned}
\frac{\mu_0(\{\overline{\Omega^{(0)}}_{0,\tau}\}_{(-A)_{\delta}})}{\mu_0(\{\overline{\Omega^{(0)}}_{0,\tau}\}_{(A)_{\delta}})} 
= \frac{\int_{\{\overline{\Omegazero}_{0,\tau}\}_{(-A)_{\delta}}}f_0(X)\dd X}{\int_{\{\overline{\Omegazero}_{0,\tau}\}_{(A)_{\delta}}}f_0(X)\dd X} 
= \exp\Bigl[-\tau[A + \epsilon(\delta, A, \tau)]\Bigr]
\end{aligned}
\end{equation}
where $\epsilon$ is a correction term obeying $|\epsilon(\delta, A, \tau)| \leq \delta$. Previous discussions of (transient albeit long-time limit) FRs in the presence of aligned static external electric and magnetic fields~\cite{barbier:2018b}, relied on the classical time-reversal and employed averages with respect to equilibrium distributions associated to opposite magnetic fields. The existence of $\mathcal{M}_{\mathrm{ext}}^{(3)}$ and $\mathcal{M}_{\mathrm{ext}}^{(4)}$, however, enables to repeat the proof of the relation in a single-system picture. The proof follows the same steps as in the standard derivation, but invokes the new operators instead of $\mathcal{M}_s$ where appropriate. In the next section, the validity of this single-system relation is illustrated \emph{via} molecular dynamics simulations.

\section{Simulations and Results}
\label{sec:results}
In the following, the theoretical results presented in the previous section are illustrated and further validated \emph{via} molecular dynamics simulations of a realistic model of liquid NaCl. The simulated system consists of 125 Na$^{+}$ and 125 Cl$^{-}$ ions in a cubic box of side 20.9\um{\angstrom}. This corresponds to a physical density $\rho = 1.3793\um{g} \um{cm}^{-3}$ (or ionic number density of 0.0275\um{\angstrom}$^{-3}$). The temperature is set to $T = 1550\um{K}$. Pair interactions are modeled using a generalized Huggins-Mayer potential, with the parameters proposed by Tosi and Fumi in Ref.~\cite{tosi:1964} and ionic charges $q_{\mathrm{Na}}=+1\um{\emph{e}}$ and $q_{\mathrm{Cl}}=-1\um{\emph{e}}$ (with \emph{e} elementary charge) for sodium and chloride, respectively. The magnetic field, directed along the $z$-axis, is set to to the value of $\vec{B} = (0, 0, 50)\um{cu}$ ($\um{cu}$ stands for code units: a detailed description of these units and of the conversion factors used in the code can be found in Ref.~\cite{mouhat:2013}), corresponding to approximately $B_z = 5\cdot10^6\um{T}$. The intensity of the field --- huge on experimental scales --- is not unusual in the context of molecular dynamics simulations of interacting systems in external fields~\cite{ciccotti:1975, ciccotti:1976,mugnai:2009,mouhat:2013,gagliardi:2016} and is dictated by the relative strength of the external to the interparticle forces. In particular, to observe appreciable effects of the external field in a reasonable simulation time, the ratio between the average interparticle forces and the average Lorentz forces has to be around one. The chosen intensity of the magnetic field results in a value of this ratio approximately equal to 0.2. Note that the magnetic field is part of the equilibrium Hamiltonian for our system. In the driven simulations, the electric field --- also directed along the $z$-axis --- is chosen to be $\vec{E} = (0, 0, 10)\um{cu}$, corresponding approximately to $E_z = 1 \cdot 10^9 \um{V}\um{m^{-1}}$. 
With this choice of the field, the ratio between the average Lorentz forces and the average electrical drift forces (absolute value) is \emph{circa}  1. 

In the simulations, periodic boundary conditions are enforced in all directions. The evolution equations~\eqref{eq:NH-EoM} are integrated \emph{via} a straightforward adaptation to the case of parallel (static and constant) magnetic and electric fields of the symplectic algorithm proposed in Ref.~\cite{mouhat:2013} for the evolution of a thermalized classical charged system in perpendicular fields. The long-range Coulombic interactions are treated using the Ewald summation method with an Ewald smearing parameter $\alpha = 0.1$ in code units. The characteristic time of the generalized Nos\'e-Hoover thermostat is fixed \emph{via} $\tau_{\mathrm{NH}} = \sqrt{Q/Gk_BT}$ with $Q = 0.1\um{cu}$.
A timestep of $\delta t = 0.25\um{fs}$ is chosen for all the simulations, ensuring that the fluctuations of the Nos\'e conserved quantity are essentially zero.

\begin{figure}[htb]
\centering
\includegraphics[width=\columnwidth]{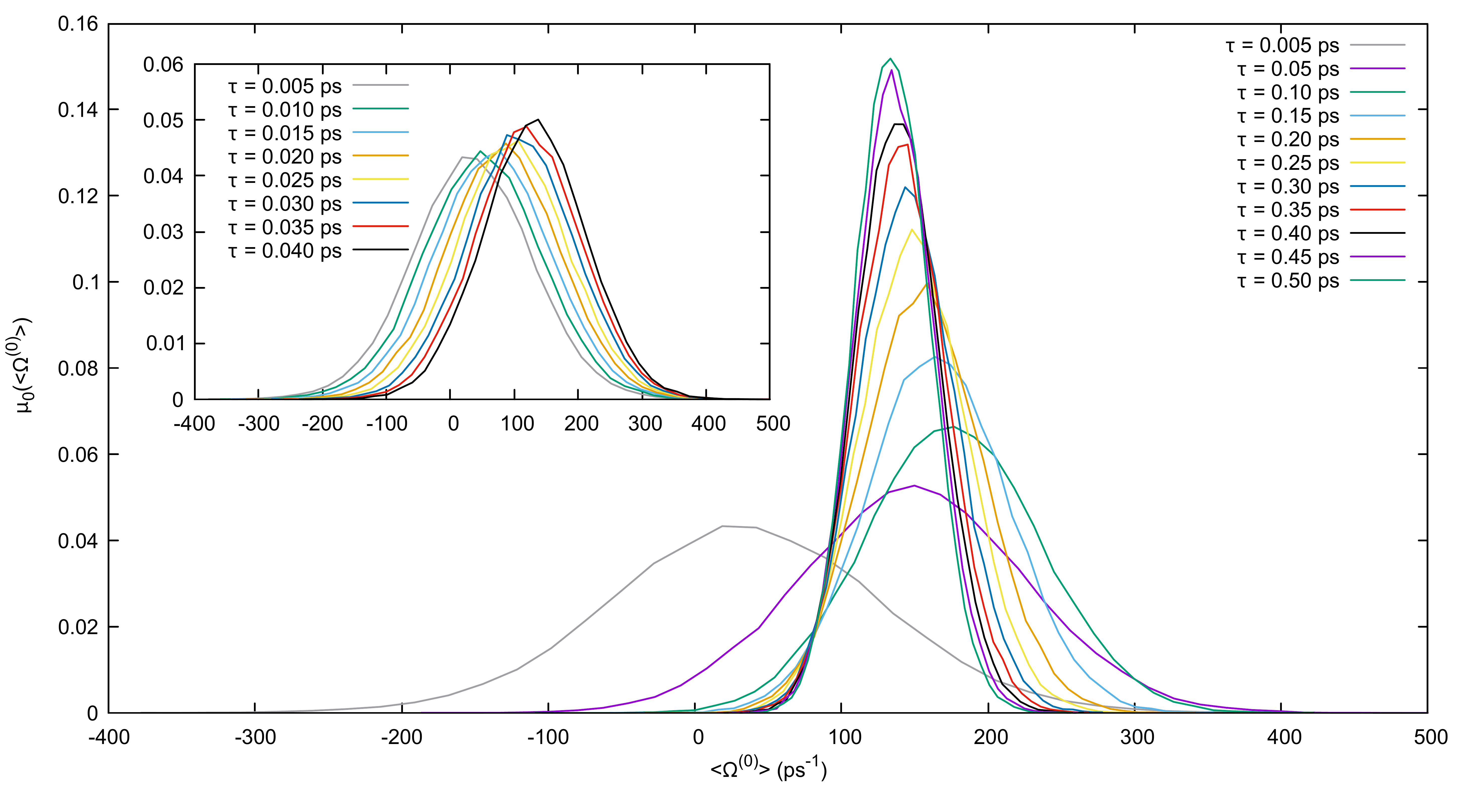}
\caption{Probability distribution functions (PDFs) estimated from the normalized histogram of the average dissipation function computed in nonequilibrium runs starting from $5\cdot10^4$ decorrelated equilibrium configurations for different values of $\tau$. The main plot shows $\tau$ ranging from $0.005$ to $0.5$\um{ps}, while the inset shows the trend for the low values of $\tau$ ranging from $0.005$ to $0.04$\um{ps}, \emph{i.e.} just after the switching on of the electric field that acts as the dissipative, nonequilibrium force.}
\label{fig:PDFs-transient}
\end{figure}

\begin{figure}[htb]
\centering
\includegraphics[width=\columnwidth]{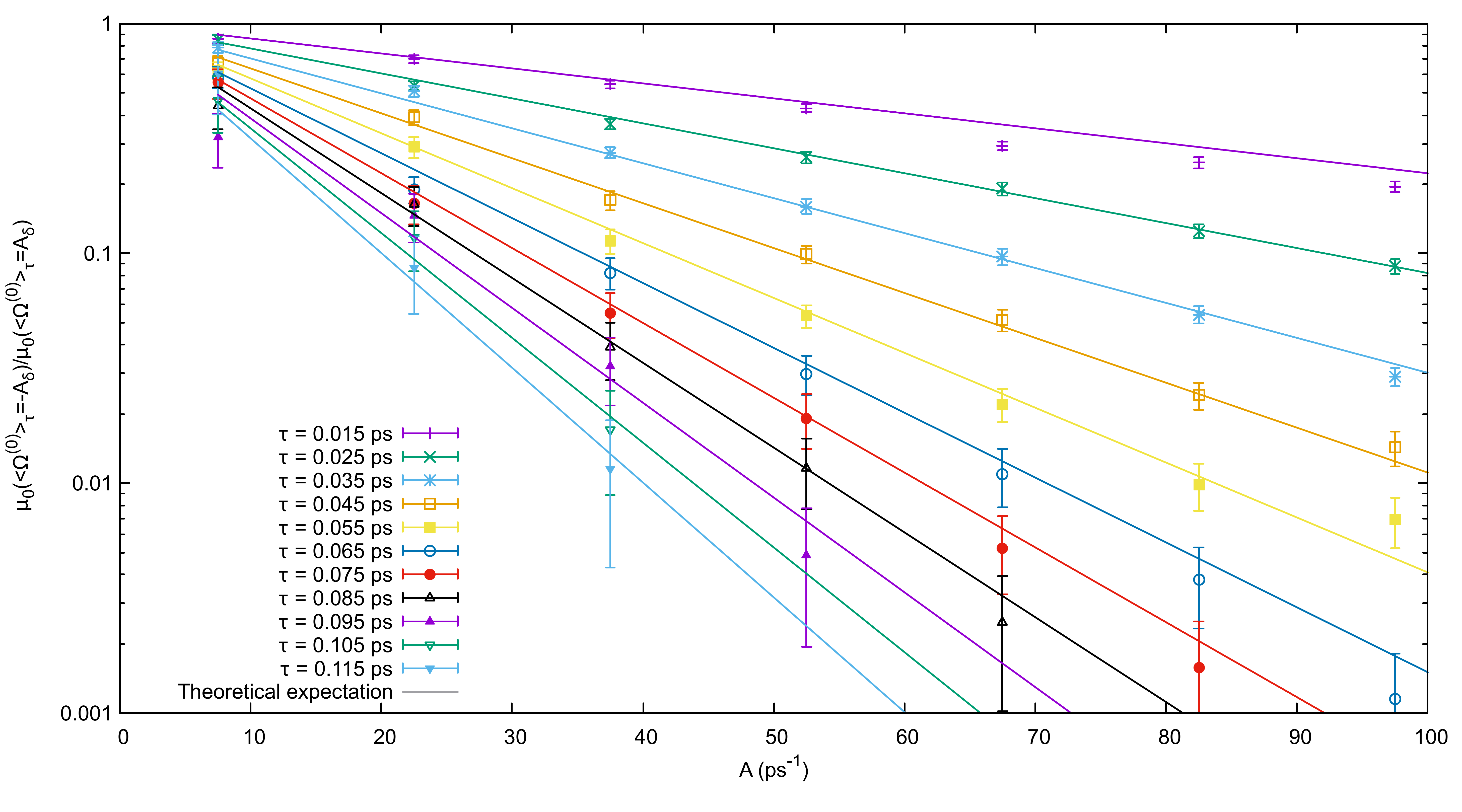}
\caption{Simulation results for the transient fluctuation relation from the simulations described in Section~\ref{sec:results}. The points represent the ratio between probabilities (on the $y$-axis) of obtaining opposite values of the average dissipation ($A$, on the $x$-axis) at different values of $\tau$ from Figure~\ref{fig:PDFs-transient}. The statistical error is obtained dividing the $5\cdot10^4$ nonequilibrium runs in 50 blocks of $10^3$ runs each and computing the normalized histogram for each of them. The error reported on the plot is obtained from the standard deviation on the single bin of the histograms relative to different blocks. Points are the simulation results, while the solid lines represent theoretical expectations. Parameters obtained form exponential fits performed on this set of numerical results are reported in Table~\ref{tab:TFR}. Note the logarithmic scale on the $y$-axis.}
\label{fig:TFR}
\end{figure}

The results discussed in this work are obtained \emph{via} the following simulation scheme. Initial conditions are fixed by placing the ions in a BCC lattice, and sampling initial velocities from the Maxwell-Boltzmann distribution corresponding to the target temperature. A preliminary equilibration run of 25\um{ps} is then executed at null electric field to enforce the target temperature \emph{via} the generalized Nos\'e-Hoover thermostat. Following this, a long equilibrium simulation ($\vec{E} = 0$) is performed to sample the equilibrium probability distribution $f_0$. In this run, phase-space configurations are sampled every 500\um{fs} (a sufficient interval to ensure decorrelation) along a trajectory of total length equal to 25\um{ns}, yielding a sample of $5\cdot10^4$ decorrelated configurations. From each of these configurations, a nonequilibrium run is started where the electric field is switched on to the reference value of $\vec{E} = (0, 0, 10)\um{cu}$. The average dissipation function, Eq.~\eqref{eq:average_Om}, is computed along each driven trajectory for a set of values of $\tau$, ranging from 5 to 500\um{fs}. 
Probability distribution functions (PDFs) for the possible values of the average dissipation at different times are then extracted through a histogramming process. 
Results for the PDFs are presented in Figure~\ref{fig:PDFs-transient}, showing the typical shifting and narrowing around the driven value of the dissipation as the simulation time lengthens.
From the probabilities of opposite values of the average dissipation functions, it is possible to check Eq.~\eqref{eq:t-FR-fin} for the system under investigation.
Results are reported in Figure~\ref{fig:TFR} for $\tau$ ranging from $0.015\um{ps}$ to $0.115\um{ps}$ together with the corresponding theoretical expectations, computed from Eq.~(\ref{eq:t-FR-fin}). As expected, the agreement between the theoretical result (solid curves) and the molecular dynamics calculation suffers as the length of the simulation and the value of $A$ increase. The exponential behavior of the calculated quantities is, however, apparent and the agreement between the two sets of data is very good. To further quantify this agreement, in Table~\ref{tab:TFR}, we show the values for $\tau$ obtained by form exponential fits performed on the numerical results and compare them with the exact value. In this case too, the agreement is very good within error bars.
\begin{table}[htb]
\caption{\label{tab:TFR}Comparison between the expected value of $\tau$ and the one obtained from the exponential fits performed on the numerical results in Figure~\ref{fig:TFR}.}
\begin{ruledtabular}
\begin{tabular}{cc}
$\tau_{\mathrm{exp}}$ (\um{ps}) & $\tau_{\mathrm{simul}}$ (\um{ps}) \\ \hline
$0.015$ & $0.0166 \pm 0.0003$ \\
$0.025$ & $0.0252 \pm 0.0003$ \\
$0.035$ & $0.0349 \pm 0.0004$ \\
$0.045$ & $0.0446 \pm 0.0005$ \\
$0.055$ & $0.0564 \pm 0.0009$ \\
$0.065$ & $0.068  \pm 0.001$  \\
$0.075$ & $0.0770 \pm 0.0008$ \\
$0.085$ & $0.086  \pm 0.003$  \\
$0.095$ & $0.097  \pm 0.008$  \\
$0.105$ & $0.104  \pm 0.006$  \\
$0.115$ & $0.112  \pm 0.008$  \\
\end{tabular}
\end{ruledtabular}
\end{table}
\section{Concluding remarks}
In this work, we have demonstrated that, and illustrated how, the transient FR can be actually verified in nonequilibrium molecular dynamics simulations of particles subject to magnetic and electric fields, and which are Nos\'e-Hoover thermostatted. The dissipation function, as well as the deterministic thermostat, have been expressed for the case in which electric and magnetic fields are parallel to each other. Although the  applicability of generalized time-reversal symmetries implicitly implies the validity of the (transient) FR, based on the Nos\'e-Hoover canonical initial equilibrium distribution, the actual possibility of verifying it in a concrete, realistic simulation is not obvious. In the first place, to the best of our knowledge, this test has never been performed in presence of a magnetic field, which substantially modifies the dynamics of particles. In the second place, a verification of the FR may be hindered by the combination of scarce statistics, and noise in the signal. Indeed, while the thermal noise becomes irrelevant at long observation (averaging) times, such long times drastically reduce the statistics of negative dissipations. Our simulations prove that the delicate balance allowing the verification of the FR, can be achieved for systems of moderately large size. Future developments will address the steady-state FR, which require further conditions to be verified, besides the time symmetry of the dynamics and of the initial phase-space probability distribution~\cite{searles:2007,rondoni:2007,searles:2013,evans:2016}.

\section*{Author contributions}
All authors have contributed equally to the conceptualization and development of the methodology and algorithm. A.C. implemented the necessary changes in an in-house code and performed the simulations. All authors contributed equally to the validation of the results and their analysis. S.B. was responsible for writing the original draft, while all authors participated equally to the reviewing and editing of the manuscript.

\begin{acknowledgments}
A.C. and L.R. have been partially supported by Ministero dell'Istruzione, dell'Universit\`{a} e della Ricerca 
(MIUR) grant ``Dipartimenti di Eccellenza 2018-2022''.
\end{acknowledgments}


\end{document}